# Atomic motion in magneto-optical double-well potentials:

# A new testing ground for quantum chaos

Shohini Ghose, Paul M. Alsing, Ivan H. Deutsch

Dept. of Physics and Astronomy, University of New Mexico, Albuquerque, NM, 87131

**Abstract:** We have identified ultra-cold atoms in magneto-optical double-well potentials as a very clean setting in which to study the quantum and classical dynamics of a nonlinear system with multiple degrees of freedom. In this system, entanglement at the quantum level and chaos at the classical level arise from nonseparable couplings between the atomic spin and its center of mass motion. The main features of the chaotic dynamics are analyzed using action-angle variables and Poincaré surfaces of section. We show that for the initial state prepared in current experiments [D. J. Haycock *et al.*, Phys. Rev. Lett. **85**, 3365 (2000)], classical and quantum expectation values show a disagreement after a finite time, and the observed experimental dynamics is consistent with quantum mechanical predictions. Furthermore, the motion corresponds to tunneling through a dynamical potential barrier. The coupling between the spin and the motional subsystems, which are very different in nature from one another, leads to new questions regarding the transition from regular quantum dynamics to chaotic classical motion.

# INTRODUCTION

Systems with multiple degrees of freedom whose constituent parts are coupled are of fundamental interest for the purpose of exploring the correspondence limit. In such cases the quantum system can explore an enormous collection of generally entangled states with no classical description. We are just beginning to characterize these entangled states at the fundamental level and realize their capabilities for information processing [1]. This disparity between the states available in the quantum and classical description is central to the mysteries of the correspondence limit. It is responsible for the distinct predictions of quantum coherent evolution and those of classical chaotic dynamics that arise in such nonlinearly coupled systems [2].

The study of quantum systems whose Hamiltonians generate classical chaos has a long history. Most studies focus on static properties ("quantum chaology" [3]) such as statistics of the energy spectrum or "scars" in the energy eigenstates [4]. As chaos is an intrinsically dynamical phenomenon, we are most interested here in understanding the time *dependent* features arising in these systems. A variety of such studies have been carried out. Most notable is the phenomenon of "dynamical localization" [5] which appears in periodically perturbed systems such as the "kicked rotor" [4]. Differences between the quantum and classical predictions for the dynamics occur due to localization of the quantum Floquet states. Dynamical localization was seen in the experiments of Raizen *et al.* [6, 7] who realized these dynamics using optical lattices – ultracold atoms in a standing wave of light. The ability to observe this phenomenon in the laboratory is evidence that the atom/optical realization provides a very clean arena in which to study coherent quantum dynamics versus nonlinear classically chaotic motion.

We have identified another nonlinear paradigm associated with trapped neutral atoms – dynamics in a magneto-optical double-potential [8]. In recent experiments by Haycock *et al.* [9], mesoscopic quantum coherence associated with the atomic dynamics has been observed. This system has some new and important features. Unlike the kicked rotor where the nonlinearity arises because of a time dependent external *classical* perturbation, in this system the nonlinear dynamics arises intrinsically from two coupled *quantum* degrees of freedom. Here, classical chaos results from the coupling between the atomic magnetic moment and its motion in the lattice. At the quantum level this leads to "entangled spinor wavepackets".

The nonlinear coupling of different degrees of freedom is often amenable to a Born-Oppenheimer approximation whereby "fast" degrees of freedoms are slaved to the "slow". Such an analysis leads to the identification of adiabatic potential surfaces. If the system strictly adheres to these surfaces, one obtains regular dynamics. The complexity arises when this approximations breaks down, which generally can occur near the anticrossings of the adiabatic potentials [10]. This leads to a variety of interesting phenomena including chaos [11-13], irreversible dissipation [14], and anomalous diffusion [15]. The latter was explored in a coupled spin-lattice system not too dissimilar from the magneto-optical potential discussed here. These analyses highlight the importance of the corrections to adiabaticity in complex dynamics. Our goal here, however, is to avoid the adiabatic approximation altogether, and instead compare the predictions of the *exact* classical dynamics to the *exact* quantum predictions. This approach is particularly useful when the system is not well described by Born-Openheimer, as is typically the case in optical lattices [16].

This article thus investigates the nonclassical nature of our dynamical system. Motional and spin degrees of freedom are of a very different nature as seen in the topology of their respective phase spaces (plane vs. sphere) and reflected in their respective Hilbert spaces (infinite vs. finite dimensional). This can lead to a disparity in the relative size of $\hbar$ in the two subsystems, raising interesting questions regarding the quantum to classical transition. In addition, for this system of entangled internal and external degrees of freedom, it is nontrivial to distinguish classically allowed from classically forbidden motion, i.e. "tunneling". The standard definition in one dimension, i.e. motion through a potential barrier, is not sufficient for systems with multiple degrees of freedom since the energy does not uniquely specify the classical trajectory. In this case, a phenomenon known as "dynamical tunneling" can occur through classically forbidden regions of phase space, which are not, however, separated by a potential barrier [17].

We have analyzed the underlying classical chaotic dynamics of our system and investigated distinct predictions of the classical versus quantum dynamics for an initial state that has been prepared in current experiments [9]. In Sec. II the physical system is briefly reviewed. Section III establishes the general predictions of the classical chaotic dynamics based on a physical picture of the primary nonlinear resonances and numerical studies via Poincaré surface of section plots. In Sec. IV. A we employ the theory of quasiprobability distributions in the coupled phase space of spin and external motion to compare and contrast classical and quantum statistics, and thereby show that the dynamics observed in the experiment are nonclassical in nature. The classical evolution diverges from the quantum dynamics much faster than on the expected logarithmic time-scale [2] and leads to a violation of the positive semidefiniteness of the density matrix [18]. Furthermore, we show that the experimentally observed nonclassical motion corresponds to tunneling through a complex region of phase space where the kinetic energy is negative (Sec. IV. B). We conclude in Sec. V with a brief discussion of further research in this area.

## II. THE MAGNETO-OPTICAL DOUBLE-WELL POTENTIAL

The physics of the magneto-optical double well has been described in previous publications [8, 19], and we summarize the salient points here. A one-dimensional optical lattice is formed by counterpropagating plane waves whose linear polarization vectors are offset at a relative angle $\theta_L$. The resulting field can be decomposed into $\sigma+$ and $\sigma-$ standing waves whose nodes are separated by $\theta_L/k$, where $k$ is the laser wave number. Atoms whose angular momenta are aligned (anti-aligned) along the lattice axis are trapped by the $\sigma+(\sigma-)$ field. A uniform magnetic field transverse to the axis would cause Larmor precession of the atom's magnetic moment, but due to the optical trap, the moment is correlated with motion of the atom between the ± wells. This correlation between spin precession and motion in the wells leads to entangled spinor wavepackets.

For the case of an atom whose electronic angular momentum is $J = 1/2$, the combined effects of the far off resonance optical potential and an applied external transverse magnetic field can be conveniently expressed in terms of a net *effective* scalar plus magnetic interaction [8],

$$U(z,\vec{\mu}) = U_J(z) - \vec{\mu} \cdot \mathbf{B}_{eff}(z). \qquad (1)$$

Here $U_J(z) = 2U_0 \cos\Theta_L \cos(2kz)$ is a scalar potential independent of the atomic moment, where $U_0$ is a constant depending on the atomic polarizability and field intensity. The effective magnetic field, $\mathbf{B}_{eff}(z) = B_x \mathbf{e}_x + B_{fict}(z)\mathbf{e}_z$, is the sum of the transverse field plus a fictitious field associated with the lattice, $\mu_B B_{fict} = -U_0 \sin\Theta_L \sin(2kz)\mathbf{e}_z$, where $\mu_B$ is the Bohr magneton. For the real alkali atoms used in experiments, the total atomic angular momentum is prepared in a hyperfine ground state with quantum number $F$. Under the circumstance that the optical trap is detuned sufficiently far from resonance so that the excited hyperfine splitting is not resolved, the form of effective potential Eq. (1) is unchanged, with the atomic magnetic moment now equal to $\vec{\mu} = \hbar\gamma\mathbf{F} = -\mu_B\mathbf{F}/F$, where $\gamma$ is the gyromagnetic ratio and $\mathbf{F}$ is the total angular momentum vector in units of $\hbar$. We consider here $^{133}$Cs, with $F = 4$, the atom used in the Jessen-group experiments [9]. The eigenvalues of the potential as a function of position result in nine adiabatic potentials, the lowest of which exhibits a lattice of double-wells.

## III. CHAOTIC CLASSICAL DYNAMICS

The Hamiltonian for the magneto-optical double-well (Eq. 1) describes the motion of a magnetic moment in a spatially inhomogeneous effective magnetic field. Generic systems of this sort have been studied in both classical and quantum circumstances, leading for example, to geometric forces [20]. An important aspect of this system is that the Heisenberg equations of motion coupling the magnetic moment to the center-of-mass dynamics are nonlinear. The corresponding classical motion is generally *chaotic*, as seen in the positive Lyapunov exponent calculated in [19], characterizing the exponential sensitivity to initial conditions. In the spin 1/2 case and for harmonic wells, we recover the Jaynes-Cummings problem [21], but without the rotating wave approximation (RWA). The classical chaotic equations of motion have been studied in quantum optics in the context of two-level atoms interacting with a single mode electromagnetic field [22], and also in condensed matter theory in the context of the small polaron problem [23]. Our system is a generalization to higher spin with no possible approximation of a single harmonic mode.

A closely related Hamiltonian was studied in the semiclassical regime by Kusnezov and coworkers [15]. Expressed in our context, their system corresponds to a spin 1/2 particle with $\theta_L$=90° (a "lin lin" lattice). Nonintegrable dynamics at the periodically distributed anticrossings that leads to anomalous diffusion over multiple anticrossings was analysed in [15].

For our system, with $\theta_L \neq 90°$ one finds that adjacent anticrossings have different energies resulting in a double-well structure (Fig. 1a). We focus on the dynamics localized to a *single* lattice site (i.e a single double-well) with *one* anticrossing bounded by high potential walls of the double-well and negligible tunneling or diffusion to neighboring sites.

We present here a more detailed analysis of the chaotic dynamics which can occur in our system. For convenience we define $\vec{\mu} = -\mu_B \mathbf{n}$, so that $\mathbf{n} \equiv \cos\theta \mathbf{e}_z + \sin\theta \left(\cos\phi \mathbf{e}_x + \sin\phi \mathbf{e}_y\right)$ is the unit direction of the atom's angular momentum, and the classical analog of $\mathbf{F}/F$. The classical equations of motion are

$$\frac{dz}{dt} = \frac{p}{m}, \quad \frac{dp}{dt} = -\frac{d}{dz}\left(U_J(z) + \mu_B \mathbf{n} \cdot \mathbf{B}_{eff}(z)\right), \quad \frac{d\mathbf{n}}{dt} = \gamma \left(\mathbf{n} \times \mathbf{B}_{eff}(z)\right), \quad (2)$$

The dynamics takes place on a four dimensional phase space $(z, p, \theta, \phi)$, which topologically is locally the tensor product of the phase plane (for the center of mass motion), and unit sphere (direction of the magnetic moment with fixed magnitude). This is equivalent to a system with two effective degrees of freedom. Nonintegrability of these equations follows since there is only one constant of the motion, the energy. The RWA would add an additional constant of the motion to the system, making the problem integrable. Without the RWA, the Hamiltonian can be made integrable under two simple physical circumstances: the case in which there is no transverse magnetic field [23], and the case of a sufficiently large transverse field so that the motion is adiabatic [24]. We consider each case separately below.

In the absence of a transverse field $(B_x = 0)$, $n_z$ becomes an additional constant of motion, which results in an integrable Hamiltonian,

$$H_0 = \frac{p^2}{2m} + C\cos(2kz + D), \tag{3}$$

$$C = U_0\sqrt{4\cos^2\theta_L + n_z^2 \sin^2\theta_L}, \quad D = \arctan(n_z \tan\theta_L/2). \tag{3a}$$

This is the Hamiltonian for a simple pendulum whose amplitude and phase depend on the constant $z$-projection of the atomic moment, as was pointed out in [15]. We present here another approach to understanding the chaos in this system using action-angle variables. The action-angle variables describing the motion of a pendulum, $(J,\psi)$, are well known to be functions of the complete elliptic integrals [25]. For energies close to the bottom of the sinusoidal potential, we can expand the elliptic integrals in a power series, keeping only the first few terms, and can therefore express $H_0$ as a function of $J$ and $\mu_z/\gamma$, which we choose to be the other action. The frequencies of precession of the corresponding angle variables $\psi$ and $\chi$ can then be computed from Hamilton's equations to be,

$$\omega_1 = \dot{\psi} = \frac{H_0}{J} = \frac{\pi}{2} \frac{\omega_0}{K(\kappa)}, \quad \omega_2 = \dot{\chi} = \gamma \frac{H_0}{\mu_z} = \frac{C \gamma H_0}{n_z \mu_B C}, \quad (4)$$

where $\omega_0 = \sqrt{4k^2|C|/m}$ is the oscillation frequency for a harmonic approximation to the sinusoidal potential, $2\kappa^2 = 1 + H_0/|C|$, and $K(\kappa)$ is the complete elliptic integral of the first kind. The frequency $\omega_1$ represents oscillation of the center of mass in the sinusoidal potential. A physical picture of the angle $\chi$ can be understood as follows. The magnetic moment precesses around the $z$ direction but at a non-constant rate since the effective field $B_z$ is changing in time. By moving to a frame that oscillates with the atom, the time dependence in the field is removed, resulting in a *constant* precession frequency $\omega_2$ about the $z$-axis. The precession angle in this frame is $\chi$. The addition of a transverse magnetic field as a small perturbation to the integrable Hamiltonian couples the oscillations of the two angles, giving rise to nonlinear resonances. The primary resonances occur when the ratio of the unperturbed frequencies is a rational number, and can be calculated for our system using Eqs. (4).

In the current experiments [9], a large transverse magnetic field is applied, which cannot be treated as a perturbation as outlined above. We therefore turn to the regime where the motion is adiabatic, and treat the nonadiabatic coupling as a perturbation. This perturbation leads to a break down of the Born-Oppenheimer approximation for our system [16]. However, sufficiently far from the hyperbolic fixed points, the system is near-integrable, allowing us to determine the resonance conditions. The integrable adiabatic Hamiltonian is obtained by setting the angle $\alpha$ between $\vec{\mu}$ and $\mathbf{B}_{eff}$ to be a constant, so that

$$H_0 = p^2/2m + U_J(z) + \mu_B |\mathbf{B}_{eff}(z)| \cos\alpha. \tag{5}$$

When $\alpha = 0$ we obtain the lowest adiabatic double-well potential (Fig. 1a). Other fixed values of $\alpha$ correspond to other adiabatic surfaces. The component of $\vec{\mu}$ along the direction of the magnetic field is now a constant of motion and serves as our new action variable. The other action of the system is obtained in the standard way by integrating the momentum over a closed orbit in the double well for a given energy and choice of the parameter $\alpha$. The precession frequencies $\omega_1$ and $\omega_2$, of the conjugate angle variables correspond respectively to the oscillation of the center of mass in the adiabatic double-well potential and precession of the magnetic moment about the local magnetic field direction in a frame oscillating with the atom as described previously. Unlike the previous case however, we cannot obtain analytical expressions for the frequencies and must resort to computing them numerically.

For the experimental parameters given in [9], Fig. 1(b) shows a Poincaré surface of section in the $p = 0$ plane and with $dp/dt > 0$, i.e. at turning points of the trajectories going from left to right. This represents a "mixed" phase space, with stable islands of periodic motion and stochastic layers at the separatrices. The primary resonance at $n_z = 0.38$ and $\phi = 0$ corresponds to a ratio of the unperturbed adiabatic frequencies of $\omega_2/\omega_1 = 4$. The nonadiabatic perturbative coupling is strong enough at these parameters to cause the previously stable primary resonance at $n_z = 0.8$ to bifurcate, and secondary resonances to appear around the points $n_z = 0.38$ and $n_z = -0.85$. The secondary resonances result from coupling between the motion around the primary islands to the unperturbed periodic motion. As the energy is increased, the primary resonances eventually disappear and global chaos sets in.

## IV. NONCLASSICAL DYNAMICS

### A. Nonclassical evolution of the quasiprobability distribution

Given the classical description of the dynamics discussed in Sec. III, we seek to determine whether the magnetization oscillations observed in [9] are truly quantum in nature. We accomplish this by calculating the dynamical evolution of the mean magnetization in a purely classical description. There are numerous approaches to a mixed quantum-classical description that have been employed, primarily by physical chemists seeking efficient numerical algorithms for describing molecular dynamics. A good summary and comparison of the various methods is discussed by Tully and coworkers [26]. Here we compute the *fully* classical evolution by first representing the initial state prepared in the experiment as a distribution of classical initial conditions for trajectories. In order to do so, we employ the theory of quasiprobability distributions on phase-space for both the external and internal degrees of freedom (analogous to the slow and fast coordinates in a Born-Oppenheimer treatment), something not typically employed in molecular dynamics [27, 28].

The state prepared at $t = 0$ was an atomic wavepacket localized on one side of the double-well potential, with a mean energy slightly above the lowest Born-Oppenheimer potential barrier (Fig.1a). The relevant representations are in terms of familiar coherent states for the motion $|\alpha = z + ip\rangle = \hat{D}(\alpha)|0\rangle$ which are translations of the harmonic oscillator ground state $|0\rangle$, and spin coherent states $|\mathbf{n}\rangle = |\xi = \theta e^{-i\phi}\rangle = \exp[(\xi \hat{J}_+ - \xi^* \hat{J}_-)/2]|-J\rangle$ for the magnetic moment, which are rotations of the spin-down state. These represent a classical direction $\mathbf{n}$ of the moment on the Bloch sphere [29]. General theories of quasiprobability distributions on the Bloch sphere have been developed analogous to those in phase space [30].

Given the initial quantum state $\hat{\rho}(0)$, we calculate the Husimi or "$Q$" quasiprobability distribution $Q(\alpha, \mathbf{n}, t = 0) = \langle \alpha|\langle \mathbf{n}|\hat{\rho}(0)|\mathbf{n}\rangle|\alpha\rangle$. We have employed the $Q$-function as it is everywhere positive and can be interpreted as a quasi-classical probability distribution. In addition, we will be interested in first order moments of observables, where issues of operator ordering that typically make $Q$ behave badly do not come into play. Phase-space distributions on the external phase space for each internal component of a two-state system have been analysed before in a semiclassical model [11-13, 27, 28]. We compute a joint Husimi distribution over both external as well as the spin phase space in order to study the dynamics on the full phase space. This four dimensional distribution function is then evolved *classically*. This was accomplished by first sampling the initial Q-distribution via a Monte-Carlo Metropolis algorithm [31], and then propagating each point in the sample according to the classical equations of motion, Eq. (3). The result gives a probability distribution at a later time, which we denote $Q_{class}(\alpha,\mathbf{n};t)$. With this function we can compute the evolution of the mean magnetization, i.e. the $z$-component of the mean angular momentum, as given in [30],

$$\left\langle \hat{F}_z \right\rangle_{class}(t) = -\sqrt{\frac{(2F+1)(F+1)(2F+2)!}{32\pi^2}} \frac{\int Q_{class}(\alpha,\mathbf{n};t)\cos(\theta)\, d^2\alpha\, d\mathbf{n}}{\int Q_{class}(\alpha,\mathbf{n};t)\, d^2\alpha\, d\mathbf{n}}. \qquad (6)$$

This result can then be compared with the quantum mechanical prediction. The quantum and classical evolutions were computed numerically using the exact Hamiltonian which already implicitly contains all non-adiabatic coupling and effective gauge potential terms [32]. The distinction between the quantum and classical dynamics are clearly shown in Fig. 2. Unlike the predictions of quantum mechanics, in the classical model the mean magnetization never becomes negative. Due to the correlation between the atomic moment and its motion in the well, an oscillation of the mean magnetization between positive and negative values corresponds to motion of the atom from one minimum of the double-well to the other. Classical dynamics thus predicts that the mean of the distribution remains localized on one side of the double-well. In contrast, the experimental data shows an oscillation between positive and negative values at a frequency well predicted by the quantum model. The only discrepancy with the ideal quantum model is that the amplitude of the experimentally observed oscillations decay due to inhomogeneous broadening in the sample [9].

A closer look at the reduced classical distribution in the phase space of position and momentum, obtained by tracing over the magnetic moment direction, shows that a part of the distribution does oscillate between wells, but the peak remains localized in one well (Fig. 3). This seems to indicate that oscillation between the wells, while not classically forbidden, is instead improbable for this distribution of initial conditions. This can been seen from the fact that the classical description of the state involves a *distribution* of energy consistent with the distribution of positions, momenta, and spin directions in the *Q*-function. High energy portions of this distribution are not classically forbidden from hopping between the left and right wells. Nonetheless, the experimentally observed oscillations of the mean atomic magnetization are much better described by the prediction of the quantum dynamics than by the corresponding classical dynamics, indicating a *nonclassical* motion of the atom between the double wells. This is not surprising given the fact that for the dynamical system and initial conditions at hand here, the actions of the system are on the order of $\hbar$.

A break between the dynamical predictions of classical and quantum theory is expected for nonlinear systems. As originally considered by Berry [2], a Hamiltonian chaotic system should exhibit observable nonclassical dynamics on a time scale *logarithmic* in $\hbar$. This follows simply by noting that in the chaotic system, the probability distribution stretches exponentially fast (set by the local Lyapunov exponent ), and develops coherence over large distances. By Liouville's theorem, the momentum distribution in the conjugate direction to the stretching is also squeezed at an exponential rate, thereby making quantum corrections to the Poisson bracket generated classical dynamics important. The time at which the chaos induced stretching of the phase space distribution causes the dynamics to depart from classical behavior is bounded from above by $t_\hbar^> = {}^{-1}\ln(I/\hbar)$ where $I$ is a characteristic action. In the limit $\hbar \to 0$, or equivalently $I/\hbar \to \infty$, classical mechanics is preserved for all times. Using a calculated Lyapunov exponent characteristic of phase space for the experimental parameters in [9], $= 1.6 \times 10^4 s^{-1}$ [33], and the smallest characteristic action of the system (here the spin), we find the time at which there is a break between classical and quantum dynamics is bounded by $t_{break} < t_\hbar^> = 89 \mu s$. As seen in Fig. 2, $t_\hbar^>$ is clearly an upper bound for the break between classical and quantum dynamics, with the true break time occurring much earlier. A more detailed analysis, identifying the scale over which the effective potential is nonlinear, is necessary to establish this time [34].

In order to quantify the nonclassical nature of these dynamics, we turn to a method recently presented by Habib *et al*. in [18]. Given an initial state $\hat{\rho}(0)$, we can compute the Wigner function through the standard Weyl transformation [35]. If we evolve this quasiprobability function for a time *t* according to the Poisson rather that Moyal bracket and then perform the inverse Weyl transformation, we obtain a "pseudo-density operator" $\hat{\rho}_{class}(t)$. An inverse Weyl-transformation on the classical propagator will not generally yield a unitary operator, and can generate nonphysical negative eigenvalues for $\hat{\rho}_{class}(t)$. This violation of the positive semidefiniteness of the pseudo-density matrix implies that the classical evolution leads to a distribution that is not a valid quantum state and has thus diverged from the quantum evolution. We have inverted the classically evolved Q-function to find the corresponding density matrix and numerically calculated its eigenvalues. This was done by first deconvolving the Q-function with Gaussian coherent states to find the Wigner function [36], and then inverting the Wigner function to obtain the corresponding density matrix. Figure 4 shows the classical eigenvalues at $t = 12.32 \mu s$. The negative eigenvalues verify that the classical evolution does violate rho-positivity. The magnitude of the rho-positivity violation is a measure of the importance of the quantum corrections to the classical evolution, and has implications for whether or not the classical limit can be recovered via decoherence [18].

### B. Tunneling

A question that remains to be answered is whether or not the experimentally observed nonclassical oscillations between the wells can be defined as tunneling. The ambiguity in the definition of tunneling in this system arises from the high dimensionality of the problem [17]. In one dimension a classical trajectory is uniquely specified by the energy, and if the potential energy is greater than this energy at any point along the trajectory, motion through this region is classically forbidden. However, for nonseparable dynamics in higher dimensions this is no longer the case since there is no longer one-to-one correspondence between energy and trajectories. In such circumstances, the phenomenon of dynamical tunneling occurs if the phase space at a fixed energy has regions bounded by separatrices. Motion between these regions is classically forbidden, but quantum mechanically the system can tunnel between them. The tunneling in this case is not defined by a potential barrier but by the classically forbidden regions of phase space. The situation becomes even more complex for nonintegrable systems, where the dynamics can be chaotic. Tunneling between two regions of phase space separated by a region of chaos can occur at a greatly enhanced rate – an occurrence known as chaos assisted tunneling [17].

In our system, the atomic spin is entangled with its motion, and thus the atom effectively moves on a higher dimensional potential surface associated with both internal and external degrees of freedom [19]. If the motion is adiabatic, then tunneling occurs when the total energy is less than the potential barrier between the adiabatic double-wells. However, for *nonadiabatic* motion, the potential barrier that defines the tunneling condition is not unique for a given energy, but depends on the trajectory of the atom on this higher dimensional potential surface as described above. Though oscillation between wells can represent quantum coherent motion, it is not obvious that this motion can be called "tunneling", especially given the finite classical probability for oscillation discussed above.

We examine first the question of adiabaticity in our system by comparing the exact energy level structure of the full Hamiltonian with that in the adiabatic approximation. In addition to the usual Born-Oppenheimer (BO) potentials $\{V_\mu(z)\}$, one must include the effect of "gauge potentials" arising due to geometric forces [32, 37]. These give corrections terms to the BO potentials in the form of an effective gauge vector and scalar field, but still within the confines of the adiabatic approximation. In the context of optical lattices, these were discussed first by Dum and Olshanii [38] and measured by the group of Raithel [39]. As discussed there, for one dimensional lattices, the vector potential vanishes and the effective scalar gauge correction to the Born-Oppenheimer potential is

$$\epsilon_\mu(z) = -\frac{\hbar^2}{2m}\langle\mu(z)|\partial_z^2|\mu(z)\rangle \tag{7}$$

where $|\mu(z)\rangle$ is the adiabatic eigenstate of the atom spin at position $z$. We solve then for the energy levels as solutions to

$$-\frac{d^2}{dz^2} + \left(V_\mu(z) + \epsilon_\mu(z)\right)\psi_\mu(z) = E\psi_\mu(z) \tag{8}$$

In Fig. 5 we plot the lowest BO potential and its gauge-corrected version. Superimposed are the energy levels as obtained from Eq. (8) and those obtained from the full Hamiltonian. It is clear that the adiabatic approximation is very coarse and does not accurately reflect the true spectrum and the resulting dynamics. For example, the energy splitting of the ground-doublet in the exact solution is 1.7 $E_R$ whereas the BO + gauge potential approximate gives 3.6 $E_R$. This calculation shows that the dynamics of our system does not follow the lowest adiabatic potential, even if we allow for gauge potential corrections to the BO potentials. The problem then is to define a tunneling condition for the non-adiabatic motion.

An unambiguous definition of tunneling is that it corresponds to motion in a classically forbidden region of phase space where the momentum must be imaginary, resulting in negative kinetic energy. In one dimension, the classical momentum at a given point is $p_{class}(z) = \sqrt{2m(E - V(z))}$, allowing us to examine the local kinetic energy. Here we can instead calculate the "kinetic energy density", so that the mean kinetic energy at time $t$ is $\langle T \rangle_t = \int T(z,t)dz$. The quantum theory gives,

$$\langle \psi(t)|\hat{T}|\psi(t)\rangle = \langle \psi(t)|(\hat{H} - \hat{V})|\psi(t)\rangle = \langle E \rangle - \int dz \sum_\mu V_\mu(z)|\psi_\mu(z,t)|^2 \qquad (9)$$

where we have expanded the wave function in the complete set of *adiabatic* eigenstates,

$$|\psi(z,t)\rangle = \sum_\mu \psi_\mu(z,t)|\mu(z)\rangle \qquad (10)$$

Thus,

$$T(z,t) = \sum_\mu \left[ \langle E \rangle - V_\mu(z) \right] P_\mu(z,t) \tag{11}$$

where $P_\mu(z,t)$ are the time-dependent populations in the BO potentials $V_\mu(z)$. The state prepared in the experiment mostly populates the lowest adiabatic potential, but at times corresponding to a Schrödinger cat-like superposition in the two wells, there is a small component in the second lowest potential due to a breakdown of the BO approximation. The mean energy $\langle E \rangle$ of this state is higher than the lowest BO potential barrier but much lower than the next adiabatic potential (Fig. 6). Thus the nonzero population in the second adiabatic state causes the kinetic energy density to be negative. The atom tunnels through a population weighted average of the two lowest BO potential barriers. The non-adiabatic transitions of the internal state thus cause the tunneling barrier to be dynamical in nature.

## V. SUMMARY AND DISCUSSION

Atoms in optical lattices provide a very clean setting in which to study dynamics arising from nonseparable couplings between two quantum subsystems which are very different in nature from one another. We have studied the chaotic dynamics for such a system and given a physical interpretation of the primary resonances. The theory of quasiprobability distributions on the tensor product of spin and motional phase space was used in order to compare the quantum and classical phase space dynamics. Our results showed that the experimental data for the atomic dynamics is best described by the prediction of quantum mechanics. Furthermore, we have clarified that this nonclassical oscillation between the wells does correspond to tunneling through a potential barrier where the kinetic energy density is negative. The important difference between tunneling in this system versus tunneling in a standard one dimensional double-well is that the barrier is not static, but depends on the evolution of the spin.

Given the disparity between the classical and quantum phase space dynamics, one can ask under what circumstances classical dynamics is recovered. One possibility is to introduce decoherence into the system. A break between the predictions of quantum and classical dynamics occurs due to rapid stretching of the chaotic phase space distribution. Decoherence acts to limit the exponential squeezing in the momentum distribution and thus diffuses the momentum uncertainty [34]. The balancing of stretching by chaos and diffusion by the environment limits the coherence length to a steady state value of $x_{coh} = x_{res}(\ /\ )^{1/2}$, where $x_{res}$ is the minimum localization length induced by the reservoir,     is the Lyapunov spreading rate, and is the damping rate. Quantum corrections to the classical Poisson bracket generated dynamics can be neglected if the wave function has spatial coherence much less than the characteristic distance in which the potential is nonlinear, $x_{NL}$, thereby recovering classical dynamics [40]. However, it has been shown that in systems that show a large violation of rho-positivity, decoherence does not succeed in recovering classical dynamics [18]. In future work we plan to explore this issue in our system through realistic models of decoherence occurring via spontaneous photon scattering. As discussed above, another intriguing aspect of our system is the intrinsic coupling between the system's internal degrees of freedom with its external motion. In some sense, the "size" of $\hbar$ for these two subsystems can be quite different. One consequence of this disparity is that decoherence can act to reduce the coherence length below the nonlinear length scale associated with one subsystem but not the other. Because these systems are entangled, an interesting question is whether the resulting dynamics can be described classically or not.

Decoherence can lead to classical behavior for *mean values* of observables [41]. However, it does not succeed in extracting localized "trajectories" from the quantum dynamics. Such trajectories are crucial for quantifying the existence of chaos both theoretically and in experiments through the quantitative measure of the Lyapunov exponents. One can recover trajectories from the quantum dynamics through the process of continuous measurements when the record is retained. Eherenfest's theorem then guarantees that well localized quantum systems effectively obey classical mechanics. The quantum "trajectories" possess the same Lyapunov exponents as the corresponding classical system [42]. The ability of a quantum measurement scheme to recover the classical dynamics increases with the size of the system action. A study of how the ratio of the internal to the external action affects the quantum-classical transition under continuous measurement of position, is currently in progress. The atom/optical system presented here provides a clean test-bed in which these issues can be explored both theoretically and in the laboratory.

**Acknowledgements:** We would like to thank the group of P.S. Jessen, as well as S. Habib, K. Jacobs, T. Bhattacharya, J. McIver, J. Grondalski and G. K. Brennen for many helpful discussions, and D. Tyler for his advice on data filtering. This work was supported by the National Science Foundation (Grant No. PHY-9732456) and the Office of Naval Research (Grant No. N00014-00-1-0575).

# References

[1] M. A. Nielsen and I. L. Chuang, *Quantum Computation and Quantum Information* (Cambridge University Press, Cambridge, 2000).

[2] M. V. Berry and N. L. Balazs, J. Phys. A: Math. Gen. **12**, 625 (1979).

[3] M. V. Berry, Proc. R. Soc. London Ser. A **413**, 183 (1987).

[4] L. E. Reichl, *The Transition to Chaos in Conservative Classical systems: Quantum Manifestations* (Springer, New York, 1992).

[5] G. Casati, *et al.*, in *Stochastic Behavior in Classical and Quantum Hamiltonian Systems*, edited by G. Casati and J. Ford (Springer, New York, 1979), Vol. 93, p. 1882.

[6] F. L. Moore, *et al.*, Phys. Rev. Lett. **73**, 2974 (1994).

[7] F. L. Moore, *et al.*, Phys. Rev. Lett. **75**, 4598 (1995).

[8] I. H. Deutsch and P. S. Jessen, Phys. Rev. A **57**, 1972 (1998).

[9] D. J. Haycock, *et al.*, Phys. Rev. Lett. **85**, 3365 (2000).

[10] D. L. Hill and J. A. Wheeler, Phys. Rev. **89**, 1102 (1953).

[11] A. Bulgac, Phys. Rev. Lett. **67**, 965 (1991).

[12] A. Bulgac and D. Kusnezov, Chaos, Solitons and Fractals **5**, 1051 (1995).

[13] H. Wagenknecht and B. Esser, Chaos, Solitons and Fractals **11**, 645 (2000).

[14] D. Kusnezov, Phys. Lett. B **319**, 381 (1993).

[15] D. Kusnezov, Phys. Rev. Lett. **72**, 1990 (1994).

[16] S. K. Dutta and G. Raithel, J. Opt. B.: Quantum and Semiclassical Optics **2**, 651 (2000).

[17] S. Tomsovic, in *Proceedings from the Institute for Nuclear Theory* (World Scientific, Singapore, 1998), Vol. 5.

[18] S. Habib, *et al.*, *The Quantum-Classical Transition in Nonlinear Dynamical Systems*, quant-ph/0010093, (2000).

[19] I. H. Deutsch, *et al.*, J. Opt. B.: Quantum and Semiclassical Optics **2**, 633 (2000).

[20] R. G. Littlejohn and S. Weigart, Phys. Rev. A **48**, 924 (1993).

[21] E. T. Jaynes and F. W. Cummings, Proc. IEEE **51**, 89 (1963).

[22] P. W. Milonni, J. R. Ackerhalt, and H. W. Galbraith, Phys. Rev. Lett. **50**, 966 (1983).

[23] D. Feinberg and J. Ranninger, Physica D **14**, 29 (1984).

[24] H. Schanz and B. Esser, Phys. Rev. A **55**, 3375 (1997).

[25] A. J. Lichtenberg and M. A. Lieberman, *Regular and Stochastic Motion* (Springer, New York, 1983).

[26] J. C. Burant and J. C. Tully, J. Chem. Phys. **112**, 6097 (2000).

[27] A. Donoso and C. C. Martens, J. Phys. Chem. **102**, 4291 (1998).

[28] C. C. Martens and J.-Y. Fang, J. Chem. Phys. **106**, 4918 (1997).

[29] F. T. Arechi, *et al.*, Phys. Rev. A **6**, 2211 (1972).

[30] G. S. Agarwal, Phys. Rev. A **24**, 2889 (1981).

[31] M. H. Kalos and P. A. Whitlock, *Monte Carlo Methods* (John Wiley & Sons Inc., New York, 1986), Vol. 1, p.73.

[32] M. V. Berry, in *Geometric Phases in Physics*, edited by A. Shapere (World Scientific, Singapore, 1989), p. 7.

[33] The Lyapunov exponent was calculated for a typical trajectory in the chaotic region of phase space at an energy equal to the mean energy of the experimentally prepared state. This represents a characteristic exponent but not necessarily the largest. A more complete calculation would average over the entire probability distribution.

[34] W. H. Zurek and J. P. Paz, Physica D **83**, 300 (1995).

[35] M. Hillery, *et al.*, Phys. Rep. **106**, 121 (1984).

[36] K. E. Cahill and R. J. Glauber, Phys. Rev. **177**, 1882 (1969).

[37] C. A. Mead, Rev. Mod. Phys. **64**, 51 (1992).

[38] R. Dum and M. Olshanii, Phys. Rev. Lett. **76**, 1788 (1996).

[39] S. K. Dutta, B. K. Teo, and G. Raithel, Phys. Rev. Lett. **83**, 1934 (1999).

[40] W. H. Zurek, Acta Phys. Pol. B **29**, 3689 (1998).

[41] S. Habib, K. Shizume, and W. H. Zurek, Phys. Rev. Lett. **80**, 4361 (1998).

[42] T. Bhattacharya, S. Habib, and K. Jacobs, eprint, quant (2000).

**Figure Captions**

FIG. 1. (a) Adiabatic potentials corresponding to the integrable hamiltonian of Eq.(5) for different values of $\alpha$. The lowest potential corresponds to $\alpha = 0$. The mean energy of the state prepared in experiments [9] is just greater than the lowest adiabatic potential barrier energy (horizontal line). The Poincaré surface of section in (b), for $p = 0$ and $dp/dt > 0$ using the parameters of [9], with $E = -186.8 E_R$ $(E_R/h = 2 kHz)$, shows the effects of the non-adiabatic perturbation term, which makes the full Hamiltonian (Eq.(1)) non-integrable.

FIG. 2. Predictions of mean magnetization dynamics. Ideal quantum theory: two-level Rabi-flopping (dashed-dotted); Ideal classical theory: localized at positive $\langle F_z \rangle$ (solid); Experimental: (circles) with a damped sinusoid fit. The upper bound on the break-time between quantum and classical dynamics is $t_\hbar^{>} = 89 \mu s$ (see text).

FIG. 3. Reduced Q–distribution in position $Q(z)$, at different times in the quantum versus classical evolution. The quantum distribution oscillates between wells while the classical distribution remains mostly on the left side, with a portion equilibrating between the wells.

FIG. 4. The eigenvalues of the classically evolved pseudo density matrix at $t = 12.32\,\mu s$. The negative eigenvalues indicate that the classical evolution violates rho-positivity and thus diverges from the quantum evolution.

FIG. 5. Lowest adiabatic potential with (dashed-dotted) and without (solid) the scalar gauge potential correction. Superimposed are the lowest two energy levels as obtained from Eq. (8) (dotted) compared to those obtained from the full Hamiltonian (dashed). The large difference between the dashed and dotted energy levels implies that the adiabatic approximation is not valid in the regime being considered in the experiment.

FIG. 6. Kinetic energy density $T(z)$ (solid curve) at (a) $t = 0$ and (b) $t = 58\,\mu s$ shown superposed on the lowest two adiabatic potentials (dotted). The mean energy of the wave function, $\langle E \rangle$, lies just above the lowest adiabatic potential $V_1(z)$, but well below the second adiabatic potential $V_2(z)$. Populations in these adiabatic states are shown in (c) and (d) at times corresponding to (a) and (b). Most of the population lies in $V_1(z)$, but the small population $P_2(z)$ in the second adiabatic state as shown in (d) causes $T(z)$ to be negative, indicating tunneling between the wells.

**Figures**

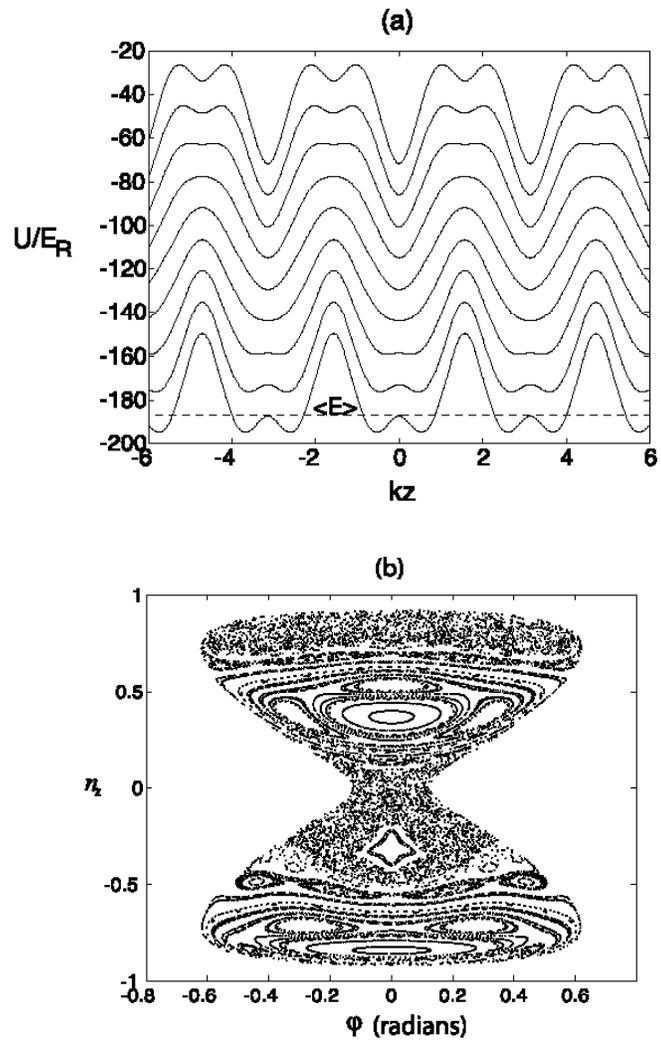

**Figure 1**

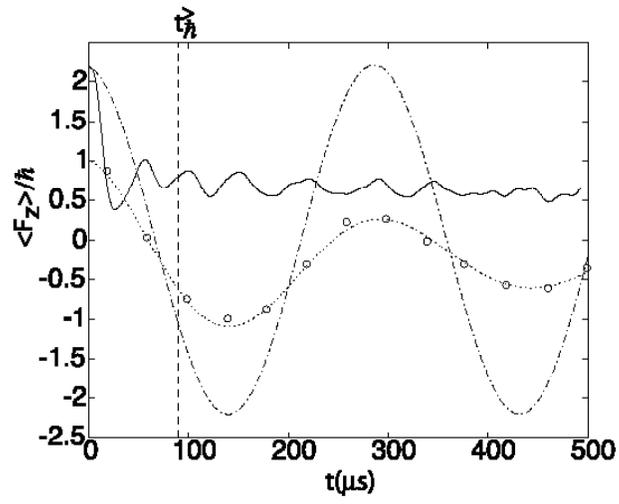

**Figure 2**

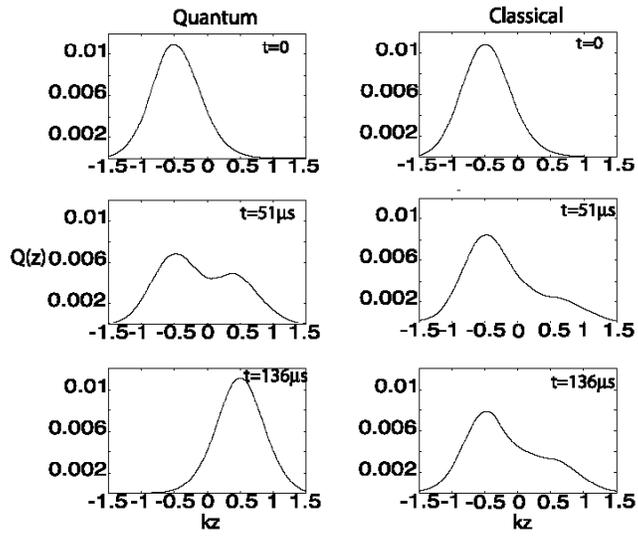

**Figure 3**

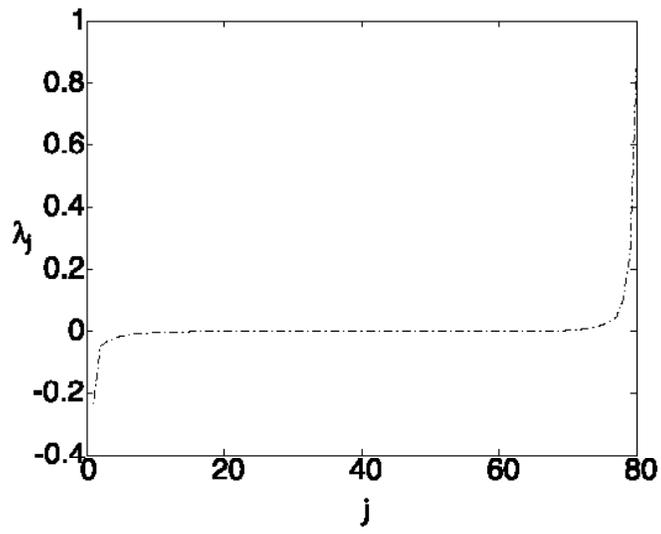

**Figure 4**

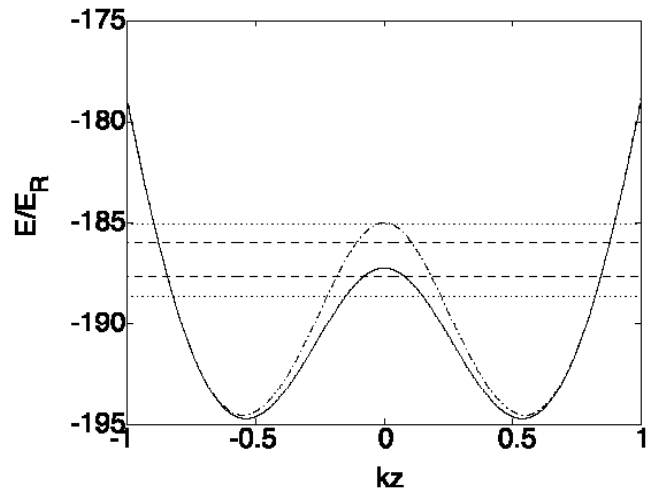

**Figure 5**

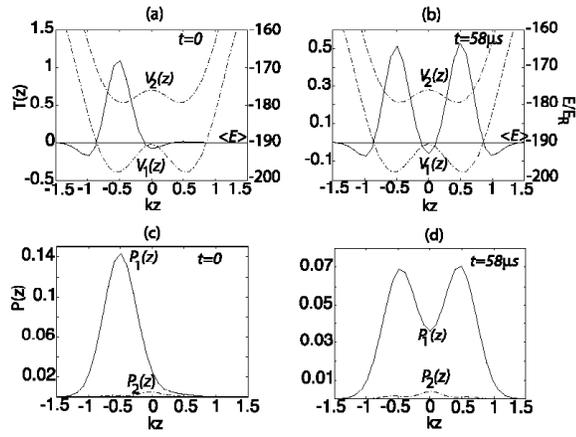

**Figure 6**